\documentclass[a4paper,fleqn]{cas-sc}

\usepackage[numbers]{natbib}
\usepackage{soul}

\usepackage{algorithm}
\usepackage{algpseudocode}
\usepackage{amsmath}
\usepackage[utf8]{inputenc}
\usepackage{subcaption}
\usepackage{graphicx}
\usepackage{pdflscape}
\usepackage{tabularx} 
\usepackage{ltablex} 
\usepackage{booktabs} 

\usepackage[utf8]{inputenc} 
\usepackage[T1]{fontenc}    

\usepackage{array}
\usepackage{tabularx, multirow, booktabs}
\usepackage{verbatim}
\usepackage{ragged2e}
\def\tsc#1{\csdef{#1}{\textsc{\lowercase{#1}}\xspace}}
\tsc{WGM}
\tsc{QE}
\tsc{EP}
\tsc{PMS}
\tsc{BEC}
\tsc{DE}

\begin{document}
\let\WriteBookmarks\relax
\def\floatpagepagefraction{1}
\def\textpagefraction{.001}
\shorttitle{Multi-Layered Security System with QKD}
\shortauthors{A. Sykot et~al.}

\title [mode = title]{Multi-Layered Security System: Integrating Quantum Key Distribution with Classical Cryptography to Enhance Steganographic Security}

\author[1]{Arman Sykot}


\affiliation[1]{organization={Department of Electrical and Computer Engineering, North South University},
                addressline={Bashundhara R/A}, 
                 city={Dhaka},
                 postcode={1229}, 
                country={Bangladesh}}

\author[1]{Md Shawmoon Azad}[orcid=0000-0001-0000-0000]
\author[1]{Wahida Rahman Tanha}
\author[1]{BM Monjur Morshed}
\author[1]{Syed Emad Uddin Shubha}[orcid=0000-0001-0000-0000]


\author[1]{M.R.C. Mahdy}[orcid=0000-0001-0000-0000]
\cormark[1]
\ead{mahdy.chowdhury@northsouth.edu}

\cortext[cor1]{Corresponding author}



\begin{abstract}
In this paper, we present a novel cryptographic system that integrates Quantum Key Distribution (QKD) with classical encryption techniques to secure steganographic images. Our approach leverages the E91 QKD protocol to generate a shared secret key between communicating parties, ensuring the highest level of security against eavesdropping through the principles of quantum mechanics. This key is then hashed using the Secure Hash Algorithm (SHA) to provide a fixed-length, high-entropy key, which is subsequently utilized in symmetric encryption. We explore the use of AES (Advanced Encryption Standard) algorithms for encrypting steganographic images, which hide sensitive information within digital images to provide an additional layer of security through obscurity.
The combination of QKD, hashing, and symmetric encryption offers a robust security framework that mitigates various attack vectors, enhancing the confidentiality and integrity of the transmitted data. Our experimental results demonstrate the feasibility and efficiency of the proposed system, highlighting its performance in terms of key generation rates, encryption/decryption speeds, and the computational overhead introduced by the hashing and steganographic processes.
By integrating quantum and classical cryptographic methods with steganography, this work provides a comprehensive security solution that is highly resistant to both quantum and classical attacks, making it suitable for applications requiring stringent security measures. This paper contributes to the ongoing research in cryptographic systems, offering insights into the practical implementation and potential benefits of hybrid quantum-classical security protocols.
\end{abstract}



\begin{keywords}
Quantum Key Distribution(QKD), 
\sep E91 Protocol, 
\sep Entanglement, 
\sep Cryptography,
\sep Advanced Encryption Standard (AES),
\sep Hashing,
\sep Secure Hash Algorithm (SHA),
\sep Encryption,
\sep Decryption,
\sep Steganography.
\end{keywords}
\maketitle

\section{Introduction}\label{sec1}
The rapid advancements in digital communication and data storage have necessitated the development of robust security mechanisms to protect sensitive information from unauthorized access and cyber threats. Traditional cryptographic methods have been the cornerstone of securing digital data; however, the emergence of quantum computing poses significant challenges to these classical techniques. Quantum algorithms like Shor's Algorithm \cite{shors} (It can successfully calculate factors of
prime numbers with exponential speedup) and Grover's Algorithm \cite{grovers} (Quantum Search Algorithm) have the potential to breach the security of classical cryptographic schemes like the RSA Algorithm\cite{milanov2009rsa}, Ellicptic-Crurve-Cryptography(ECC)\cite{hankerson2021elliptic} and many other. Shor's Algorithm takes asymptotically ${\displaystyle O\!\left((\log N)^{2}(\log \log N)(\log \log \log N)\right)}$ steps on a quantum computer. And Grover's quantum algorithm for unstructured search that finds with high probability the unique input to a black box function that produces a particular output value, using just ${\displaystyle O({\sqrt {N}})}$ evaluations of the function due to the principle of quantum mechanics, where ${\displaystyle N}$ is the size of the function's domain. 
Quantum Key Distribution (QKD)\cite{alleaume2014using} presents a promising solution by leveraging the principles of quantum mechanics to enable secure key exchange, offering unparalleled security guarantees against eavesdropping.
The E91 QKD \cite{e91main} protocol, introduced by Ekert in 1991, is based on the phenomenon of quantum entanglement and provides a means for two parties to generate a shared secret key with absolute security assured by the laws of physics. Where two parties creates a Bell State\cite{gisin1998bell}, such as $\psi_{\pm}=\frac{1}{\sqrt{2}}|01\rangle \pm |10\rangle$. Despite its theoretical advantages, the practical implementation of QKD systems requires seamless integration with existing cryptographic infrastructures to ensure comprehensive data protection.
\begin{figure}[hbt!]
    \centering
    \includegraphics[width=1\linewidth]{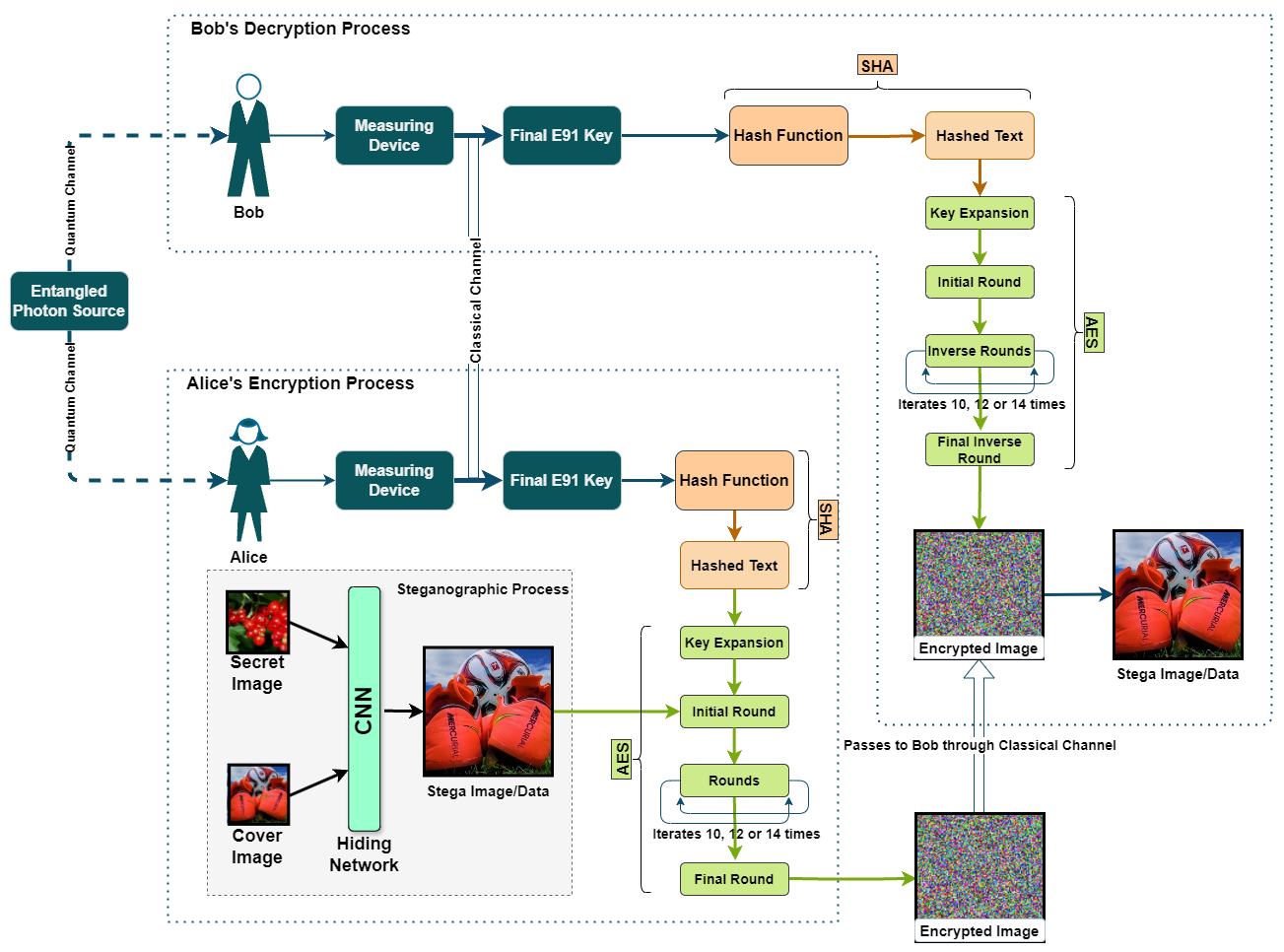}
    \caption{A high-level overview of the proposed architecture for the Multi-Layered Security System. E91 protocol shared secret keys between Alice and Bob from an entangled photon source after doing some bases measurements described in section \ref{e91 protocol}. Then the key passes to the Secure Hash Algorithm function to create a hashed key of the secret shared key. Then, it passes to the AES encryption algorithm to encrypt the Steganographic Data, which we described in \ref{deepstega}. Thus we created a robust security system to enhance the cryptographic method and steganographic security}
    \label{Figure: qkd-aes-hash}
\end{figure}
In parallel, steganography has emerged as an effective technique for concealing information within digital media, such as images, audio, and video files. By embedding data in a way that is imperceptible to the human eye, steganography adds an additional layer of security through obscurity. However, the security of steganographic methods heavily relies on the robustness of the underlying cryptographic algorithms used for data encryption\cite{ozkaynak2018brief}.\\
This paper presents a novel approach that combines the strengths of QKD, classical encryption, and steganography to enhance the security of digital communication. Our system utilizes the E91 protocol to generate a highly secure shared key, which is then hashed using the Secure Hash Algorithm (SHA)\cite{eastlake2001us} to ensure a fixed-length, high-entropy key suitable for symmetric encryption. We explore the use of Advanced Encryption Standard (AES) \cite{aesrij} algorithms to secure steganographic images, thereby protecting hidden information with a strong cryptographic framework.
The integration of QKD with classical encryption techniques addresses several security challenges, including resistance to both quantum and classical attacks. By combining these advanced cryptographic methods with steganography, our approach not only ensures the confidentiality and integrity of the transmitted data but also leverages the concealment properties of steganography to provide an additional layer of security.\\
In this paper, we provide a detailed description of our system architecture, including the key generation process using the E91 QKD protocol, the hashing mechanism, and the encryption of steganographic images. We conduct a comprehensive security analysis to evaluate the resilience of our system against various attack vectors and perform extensive performance evaluations to assess the efficiency of our approach. Our results demonstrate that the proposed system offers a robust and efficient solution for secure digital communication, making it suitable for applications that require stringent security measures.
By integrating quantum and classical cryptographic techniques with steganography, this work contributes to the ongoing research in cryptographic systems and provides valuable insights into the practical implementation and benefits of hybrid quantum-classical security protocols. This novel approach paves the way for future developments in secure communication technologies, offering enhanced protection against the evolving landscape of cyber threats.

\section{Literature Review}\label{litrRev}
Quantum Key Distribution (QKD) combined with classical cryptography presents a promising frontier for enhancing steganographic security. Here, we will explore the state-of-the-art research and methodologies that have integrated QKD with classical cryptographic techniques and their application in steganography, providing an extensive overview of recent advancements and their implications. 

Classical cryptography refers to encryption techniques that rely on mathematical algorithms and computational complexity to secure data. It is based on established mathematical principles and has been the foundation of secure communication for decades. Classical cryptographic methods are divided into different algorithms such as Symmetric key algorithms- Advanced Encryption System(AES), Data Encryption Standard (DES), Asymmetric key algorithms- Rivest–Shamir–Adleman(RSA), Elliptic Curve Cryptography (ECC) and Hash Functions.
Among these, ECC enhances the security of open communication networks. A recent survey investigates scientific concepts, innovative methodologies, and implementations of ECC. ECC-based schemes are considered to be more secure for cloud computing, e-health, and e-voting than RSA and Diffie–Hellman schemes. It is also proven to be more advanced in distributed and asynchronous networking, providing significant security advantages over RSA and Diffie–Hellman.\cite{ULLAH2023100530}
An ECC-based novel image encryption methodology has been presented for grayscale and colour images. It employs a combination of primitive polynomials, invertible Suslin matrices, and logistic chaotic maps. This scheme targets to enhance security and efficiency in image encryption, making it robust against various attacks and suitable for secure image transmission.\cite{Sharma2024}

Another chaotic-based image encryption scheme combines Arnold's cat map for pixel shuffling, elliptic curve cryptography (ECC) for pixel value encryption using public and private keys, and a genetic algorithm to optimize key generation. This integration enhances data security by introducing chaos, robust encryption, and optimized key management. The approach indicates strong resistance against statistical attacks. Additionally, the extensive key space enhances resilience against brute force attacks.\cite{Kumar2024}

QKD uses the principles of quantum mechanics to exchange the key, which is theoretically assumed to be secure even for quantum hackers and other more sophisticated attackers. Some research works have discussed the enhancement of QKD with other classical cryptography algorithms for practicality in defence. An approach of incorporating QKD into hybrid quantum-classical networks highlighted some methods to integrate QKD into existing Transport Layer Security (TLS) protocols, enhancing security for data transmission over optical fibres. It used Post Quantum Cryptography (PQC) algorithms to secure the delivery of QKD keys, demonstrating the complementing strengths of QKD and PQC technologies.\cite{10.1007/978-3-031-41181-6_42}. 
Another comprehensive security framework has been presented to ensure the security of QKD protocols. It addresses finite-size security, key generation, error correction, privacy amplification, and parameter estimation. Alice and Bob exchange quantum states and then publicly communicate to reconcile their keys, ensuring minimal errors and a high level of secrecy. Parameter estimation calculates the error rates and other statistical measures to ensure the security of the key. Then the shared keys are distilled into highly secure keys that are resistant to potential eavesdropping by hashing through two universal hash functions.\cite{arXiv:2305.05930}

Cryptographic hash functions are essential for ensuring data integrity and authentication. SHA-256, part of the SHA-2 family, is widely used in various security protocols and applications. It produces a 256-bit hash value from input data, providing strong collision resistance and preimage resistance and, therefore, widely used in encryption methods.
An approach integrated DNA sequence operations with a chaotic system (specifically the Lorenz system) and the Secure Hash Algorithm SHA-2 to enhance the security of image encryption. SHA-2 adds an extra layer of security and integrity to the encrypted images.\cite{10.1007/s11071-015-2392-7}
Another study proposed a design of a colour image encryption technique performed on DNA sequences with the objective of making the design plaintext-sensitive. This scheme integrates 1D chaotic systems with the SHA-256 hash function of a plain image's RGB components to dynamically update secret keys. This process enhances security by making keys sensitive to any changes in the plain image, mitigating various plaintext attacks like chosen plaintext and known plaintext attacks. The scheme transforms real numbers into integers, converting key matrices derived from random integer sequences into DNA sequences. It employs two different diffusion mechanisms and adjusts the permutation sequence with changes in the secret key's block size. Experimental results show that the scheme is sensitive to variations in the plaintext and secret key parameters, which justifies its robustness.\cite{8940558}

Steganography is the art of hidden communication over a
traditional route. QKD, embedded with this security measure, introduced a new way of fortifying data. A security model visualized the idea of using the Bernstein-Vazirani algorithm for integrating data into a quantum circuit and QKD protocol BB84  to obtain a secret key, which ensures that only authorized parties can access the quantum message. 
This quantum steganography protocol increases the hidden channel capacity. It also withstands intercept-resend and auxiliary particle attacks due to the randomness in the distribution of information and secret messages. Also, it uses more Bell states to reduce the detectability of hidden messages.\cite{Yalla2022}. 
Another study presented a novel protocol for quantum steganography using discrete modulation continuous variable QKD to address vulnerabilities in existing protocols. By incorporating reverse communication, the proposed protocol avoids detection by steganalysis(a procedure that focuses on detecting hidden information or data within various types of digital media), ensuring the security and secrecy of the hidden messages.\cite{Joshi2022}

Our work comprises multi-layer security of QKD protocol with classical cryptographic encryption and hash function, all embedded together to conceal steganographic data to a more secure extent.

\section{Proposed Architecture}\label{proposal}
Our proposed architecture integrates quantum key distribution (QKD) E91 protocol with classical cryptographic methods and deep steganography to ensure a highly secure image data encryption system. This system leverages the E91 protocol for QKD to generate cryptographic keys, Secure Hash Algorithm (SHA) to hide and make fixed key length, AES for encryption, and deep learning models for steganography. The architecture ensures robust data security through a combination of quantum and classical techniques. The Figure: \ref{Figure: qkd-aes-hash} gives a high-level overview of our Multi-Layered Security System. 

\subsection{Initialization}\label{initial}
\begin{figure}[hbt!]
    \centering
    \includegraphics[width=0.35\linewidth]{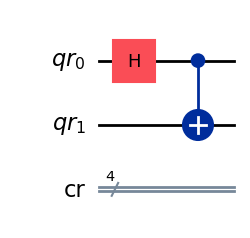}
    \caption{Singlet State Circuits}
    \label{fig:singlet}
\end{figure}

We have created singlet circuit to create entangled qubits using Qiskit SDK\cite{adedoyin2018quantum} for quantum computing. So that they can make secured shared key through E91 quantum key distribution protocol which is described in section \ref{e91 protocol}. We added a Hadamard(H) gate to make the qubit in superposition and added a CNOT gate to control the $2^{nd}$ qubit. We have explained these quantum logic gates in section \ref{quantum logic gate}. Figure \ref{fig:singlet} shows the initial singlet state circuits of E91 protocol.

\subsection{Measurement}\label{measureResult}
\begin{figure}[h!]
    \centering
    \includegraphics[width=0.85\linewidth]{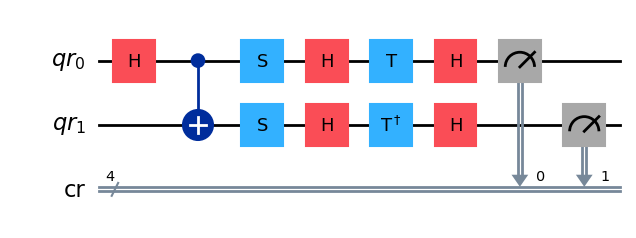}
    \caption{Measurement circuit in different bases like $Z$, $\frac{(X+Z)}{\sqrt{2}}$, $X$ and $\frac{(X-Z)}{\sqrt{2}}$}
    \label{fig:measure}
\end{figure}

Alice and Bob select some bases independently which is described in section \ref{measurement}. Then, they make independent measurements based on those bases. They communicate through classical channels to verify their same measurement bases and generate secret keys from those measurement bases. Those measurement bases are not the same for each other; they prepare those bases for the CHSH test to check the eavesdropper, which is described in section \ref{CHSH}. Figure: \ref{fig:measure} shows what a measurement circuit looks like on Qiskit SDK for quantum computing.

\subsection{Hash Function}\label{hashalgo}
After extracting the shared secret key from E91 protocol, Alice and Bob creates a hash text using Secured Hash Algorithm(SHA). SHA-256 is described in section \ref{SHA}. Hashing helps to hide the original length of the secret key and creates a fixed length. So if somehow an eavesdropper gets access to the hash text, he/she will still be unable to extract the original key.\\ 

For 500 singlet state, we will get one of the following keys:
\\$01000100101010001010111001010001000010011110010001110011010011011001111000111111101000101100000000$\\
From section \ref{SHA} equation no. \ref{hasheqn}, we get the hash value of the following key as:\\
$b0c1de302023c06df2ca56b7206da959dccc9cc8d25f983f921b51664a7e2851$\\

Similarly, for 250 singlet state, we will get one of the following keys:\\ $0110110001111001010101101110000101000011101100100$\\
The hash value of the following key is:\\
$6c26412c58131663b4034e37bf71318c115589186085671c27f25b368e4cbb4e$\\

Similarly, for 100 singlet state, we will get one of the following keys:\\
$1010011110001000111011$\\
The hash value of the following key is:\\
$dd116ea845b69b000cfde2831b67e5ac53544fd9688b86123fef6235e34af651$\\

Similarly, for 25 singlet state, we will get one of the following keys:\\ 
$10010111$\\
The hash value of the following key is:\\
$459c2daec5458568864215c57d12fa0ae28243b080971bb90d09fe020f8f265e$\\

So, we can write, 
\begin{align}\label{hashedkey}
    H = H'(E91_{key})
\end{align}
Here $H$ is the Hashed text, $H'()$ is the Hash function, and $E91_{key}$ is the shared secret key generated from the E91 protocol.

\subsection{Encryption and Decryption using AES}\label{encryptdecryptAES}
\begin{figure}[hbt!]
    \centering
    \includegraphics[width=0.65\linewidth]{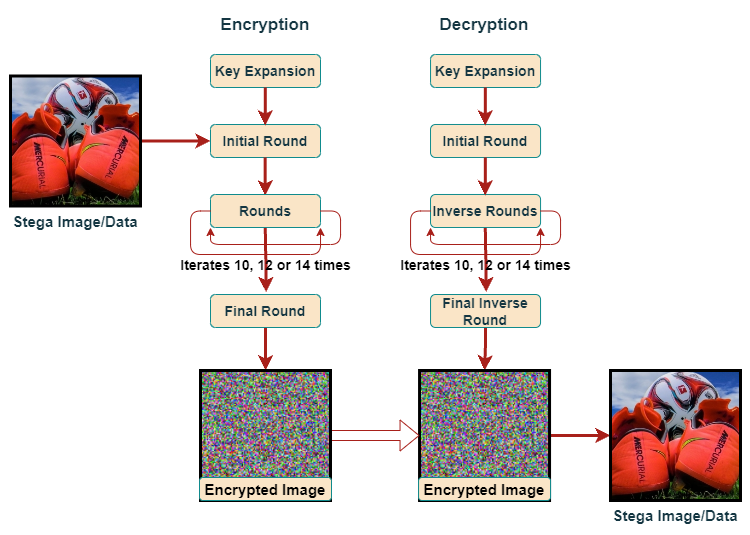}
    \caption{An overview of AES encryption and decryption method}
    \label{fig:aes}
\end{figure}

After generating the shared secret key from E91 protocol and using the Hash function we created a cryptographic hashed key $H$ on equation no. \ref{hashedkey}. This hashed key $H$ enters the Key Expansion section of AES encryption and then encrypts the desired data. We have described all the methods in section \ref{AES}. The figure: \ref{fig:aes} shows the detailed encryption and decryption process of data. \\
So, after encrypting data we can write,

\begin{align}
    C = AES_{encrypt}(I, H'(E91_{key})) 
\end{align}

from equation no. \ref{hashedkey}, we got, $H = H'(E91_{key})$\\
So, we can write, 

\begin{align}
    C = AES_{encrypt}(I, H)
\end{align}

Here, $C$ = Cipher Data, $AES_{encrypt}()$ is the encryption function, $I$ is the stega image/plain data, $H$ is the hashed key. 
Similarly, for decryption,
\begin{align}
    I = AES_{decrypt}(C, H)
\end{align}
Here, $AES_{decrypt}()$ is the decryption function.

\subsection{Deep Learning Based Steganography}
\begin{figure}[hbt!]
    \centering
    \includegraphics[width=0.85\linewidth]{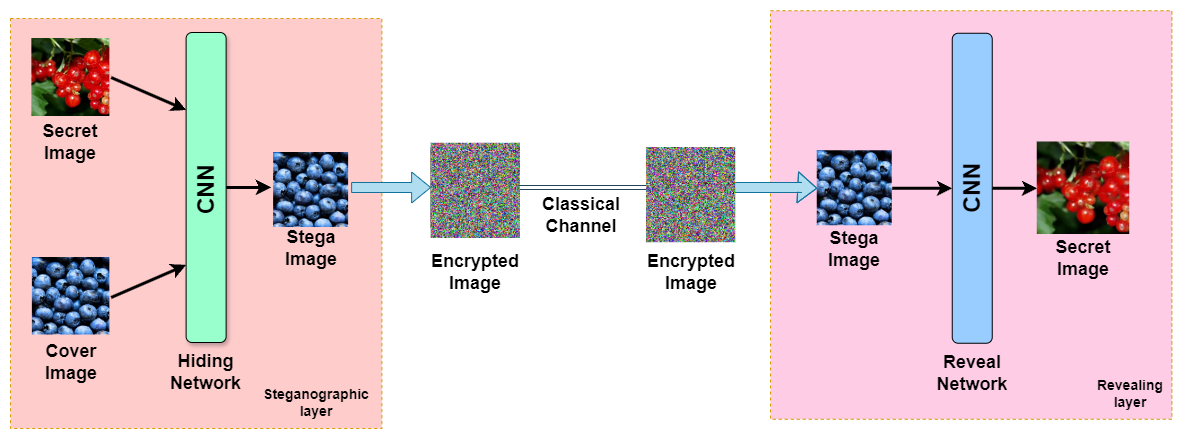}
    \caption{An overview of deep-learning based steganography, where a secret and cover image is being embedded by CNN, which is the hiding network. Thus, a Stega Image is created. Then we used our novel approach to encrypt the stega image and passed it through a classical channel and decrypted it. Another CNN, which is the revealing network, is used to unhide the secret image.}
    \label{fig:deepstega}
\end{figure}

In our proposed method, we have shown a novel cryptographic method to encrypt steganography data. We could have shown any type of data. Since steganography itself is a cryptographic technique, we choose to encrypt a cryptographic technique. In section \ref{deepstega}, we talked about deep learning-based steganography, which we used to embed a secret and a cover image. Figure: \ref{fig:deepstega} shows how deep steganography works.

\section{Result Analysis}\label{result}
We have used Qiskit SDK for Quantum Computing by IBM to generate different results like key generation, encryption rate, entropy analysis and differential attack analysis. We have also shown different histogram results of an encryption and decryption image. 

\subsection{Entropy analysis of encrypted images}
The Shannon entropy\cite{6773024} of a cipher is a measure of its randomness and unpredictability, which are crucial properties for secure encryption. High entropy means that the encrypted data appears random and has no discernible patterns, making it difficult for attackers to predict or deduce the original data.
\begin{align}\label{entropyeqn}
    H(x) = -\sum^n_{i=1} P(x_i)log_b(P(x_i))
\end{align}
Here, $H(x)$ is the entropy of the random variable $x$. $P(x_i)$ is the probability outcome.
An encryption technique should ideally output an encrypted image with an entropy value of 8, which is computed using equation no.\ref{entropyeqn}. Table: \ref{tab:entropy} shows the entropy value of different images of different pixels. We have tested 4 different sizes of pixels of many images, and we have always found the value of entropy $\approx 8$.

\begin{table}[hbt!]
    \centering
    \caption{Entrophy computation of encrypted images of different pixel size.}
    \label{tab:entropy}
    \begin{tabular}{cc}\hline
        Pixel Size & Entropy cipher\\
        \hline
        64 x 64 & 7.9842\\
        \hline
        128 x 128 & 7.9894\\
        \hline
        256 x 256 & 7.9986\\
        \hline
        512 x 512 & 7.9995\\
        \hline
    \end{tabular}
\end{table}

\subsection{NPCR and UACI attack analysis}
NPCR\cite{hussain2013change} (Number of Pixels Change Rate) and UACI\cite{saidi2020number} (Unified Average Changing Intensity) are metrics used to evaluate the effectiveness of image encryption algorithms by measuring the difference between an original image and its encrypted version.\\
NPCR measures the percentage of different pixel values between the original and encrypted images. It indicates how many pixels have changed due to the encryption process. The formula for NPCR is:
\begin{align}
    NPCR = \frac{\sum_{i,j}D(i,j)}{M \times N} \times 100\% 
\end{align}
where, $D(i, j)$ is a binary array where $D(i, j) = 1$ if pixel value at position $(i, j)$ in the original image is different from the encrypted image, and $D(i, j) = 0$ otherwise. $M \text{ and } N$ are the dimensions of the image.\\
UACI measures the average intensity of differences between the original and encrypted images. It quantifies the average intensity of the differences in pixel values.The formula for UACI is:
\begin{align}
    UACI = \frac{1}{M \times N}\sum_{i, j} \frac{|C_1(i,j)-C_2(i,j)|}{255} \times 100\%
\end{align}
where, $C_1(i,j)$ and $C_2(i,j)$ are the pixel values at position $(i, j)$ in the original and encrypted images, respectively and $D(i, j) = 0$ otherwise. $M \text{ and } N$ are the dimensions of the image. Table: \ref{tab:ncpruaci} shows differential attack analysis.
\begin{table}[hbt!]
    \centering
    \caption{Computations of differential attack analysis.}
    \label{tab:ncpruaci}
    \begin{tabular}{ccc}\hline
        Pixel Size & NCPR(\%) & UACI(\%)\\
        \hline
        64 x 64 & 99.6337 & 54.4415\\
        \hline
        128 x 128 & 99.8229 & 57.7351\\
        \hline
        256 x 256 & 99.5432 & 55.4843\\
        \hline
        512 x 512 & 99.7717 & 56.9590\\
        \hline
    \end{tabular}
\end{table}

\subsection{Key Generation Rate of E91 Protocol}
Here we generated key generation rate and time for the E91 protocol by using Qiskit SDK for Quantum Computing. Table: \ref{tab:keygen} shows the different rates and times in seconds and bits per second.  
\begin{table}[hbt!]
    \centering
    \caption{Key Generation Rate}
    \label{tab:keygen}
    \begin{tabular}{cccc}\hline
        Singlet State Used & Key Length & Key Generation Time(s) & Key Generation Rate(bps)\\
        \hline
        25 & 7 & 4.39 & 1.59\\
        \hline
        100 & 25 & 4.66 & 5.37\\
        \hline
        250 & 57 & 5.42 & 10.52\\
        \hline
        500 & 106 & 8.66 & 12.24\\
        \hline
    \end{tabular}
\end{table}

\subsection{Encryption, Decryption Rate}
Here we will see the encryption and decryption\cite{al2007simple} time rate of our proposed system for different image pixel sizes. Table: \ref{tab:enrypdecryp} shows the results of execution time complexity.
\begin{table}[hbt!]
    \centering
    \caption{Execution time complexity of the proposed system}
    \label{tab:enrypdecryp}
    \begin{tabular}{ccc}\hline
        Pixel Size & Encryption Time(s) & Decryption Time(s)\\
        \hline
        64 x 64 & 0.00098 & 0.00100\\
        \hline
        128 x 128 & 0.00100 & 0.00001\\
        \hline
        256 x 256 & 0.00100 & 0.00200\\
        \hline
        512 x 512 & 0.00300 & 0.00399\\
        \hline
        \textbf{Average} & \textbf{0.00149} & \textbf{0.00175}\\
        \hline
    \end{tabular}
\end{table}

\subsection{Histogram Analysis}
\begin{figure}\label{img1}
	\centering
	\begin{subfigure}{0.4\linewidth}
		\includegraphics[width=\linewidth]{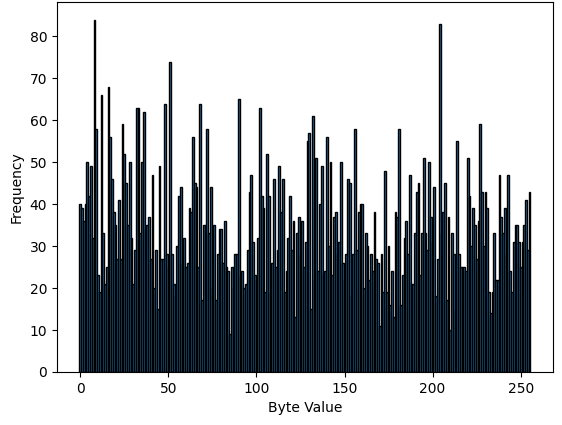}
		\caption{Original Image}
		\label{fig:subfig1A}
	\end{subfigure}
	\begin{subfigure}{0.4\linewidth}
		\includegraphics[width=\linewidth]{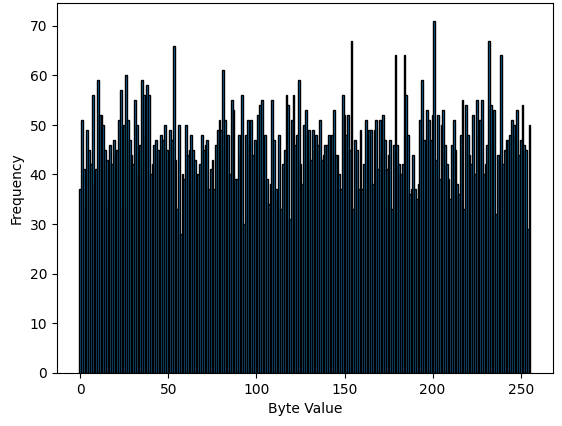}
		\caption{Encrypted Image}
		\label{fig:subfig1B}
	\end{subfigure}
    \caption{Original and Encrypted Image histogram of 64x64 pixel image}
\end{figure}

\begin{figure}\label{img2}
	\centering
	\begin{subfigure}{0.4\linewidth}
		\includegraphics[width=\linewidth]{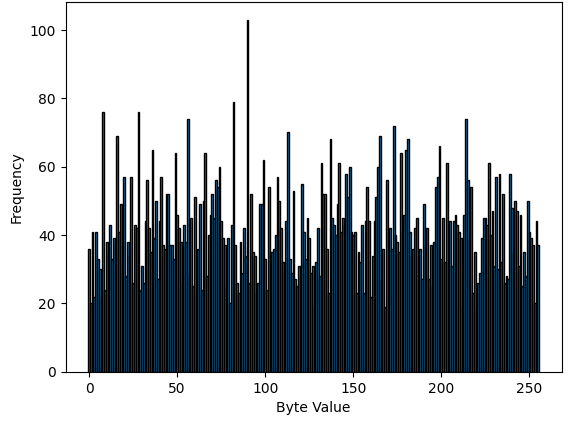}
		\caption{Original Image}
		\label{fig:subfig2A}
	\end{subfigure}
	\begin{subfigure}{0.4\linewidth}
		\includegraphics[width=\linewidth]{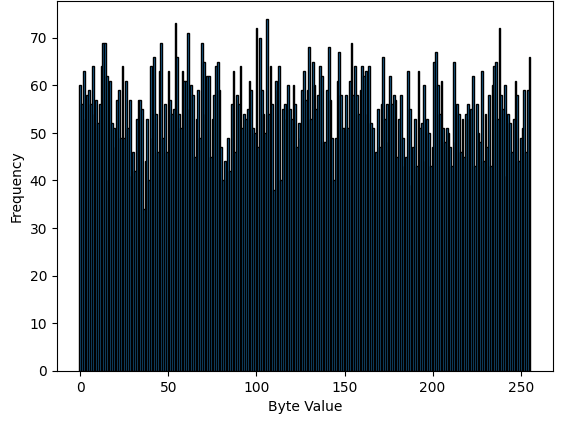}
		\caption{Encrypted Image}
		\label{fig:subfig2B}
	\end{subfigure}
 \caption{Original and Encrypted Image histogram of 128x128 pixel image}
\end{figure}

\begin{figure}\label{img3}
	\centering
	\begin{subfigure}{0.4\linewidth}
		\includegraphics[width=\linewidth]{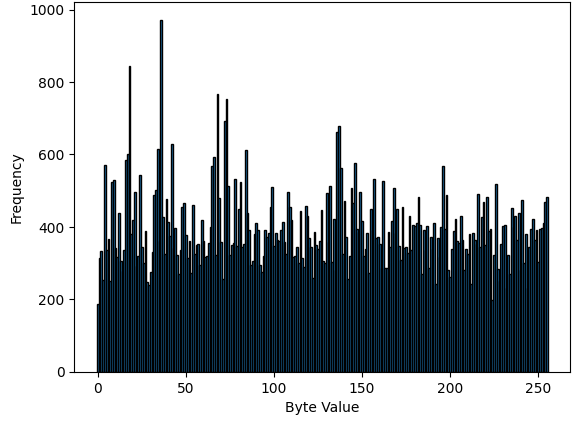}
		\caption{Original Image}
		\label{fig:subfig3A}
	\end{subfigure}
	\begin{subfigure}{0.4\linewidth}
		\includegraphics[width=\linewidth]{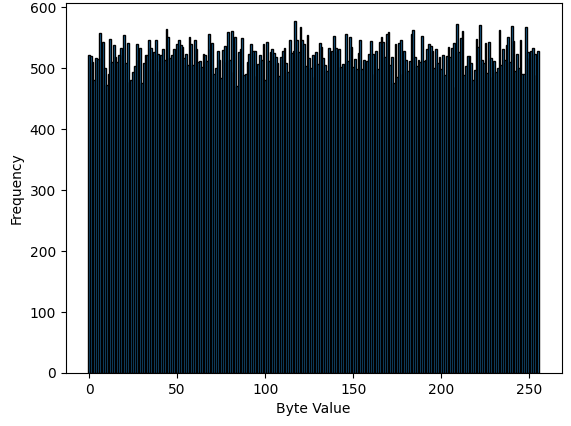}
		\caption{Encrypted Image}
		\label{fig:subfig3B}
	\end{subfigure}
 \caption{Original and Encrypted Image histogram of 256x256 pixel image}
\end{figure}

\begin{figure}\label{img4}
	\centering
	\begin{subfigure}{0.4\linewidth}
		\includegraphics[width=\linewidth]{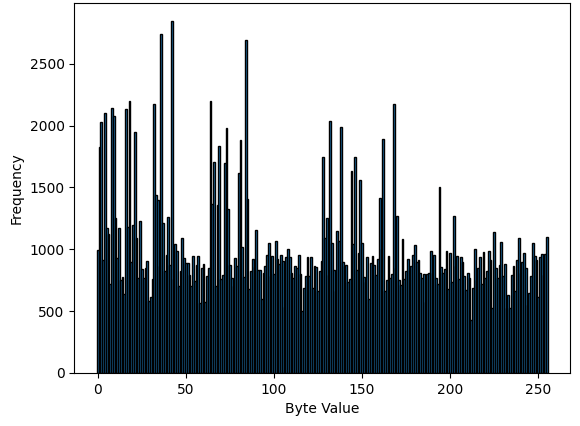}
		\caption{Original Image}
		\label{fig:subfig4A}
	\end{subfigure}
	\begin{subfigure}{0.4\linewidth}
		\includegraphics[width=\linewidth]{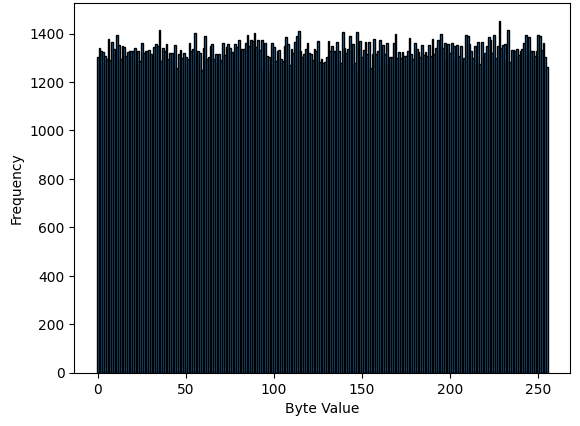}
		\caption{Encrypted Image}
		\label{fig:subfig4B}
	\end{subfigure}
 \caption{Original and Encrypted Image histogram of 512x512 pixel image}
\end{figure}

In figure: \ref{fig:subfig1A} \ref{fig:subfig1B}, figure: \ref{fig:subfig2A} \ref{fig:subfig2B}, figure: \ref{fig:subfig3A} \ref{fig:subfig3B}, and figure: \ref{fig:subfig4A} \ref{fig:subfig4B}, we have shown the different histograms\cite{scott2010histogram} of different pixel images, and we can see how the histogram changes when the image is being encrypted. 

\subsection{Key-Sensitivity Analysis}
\begin{figure}
	\centering
	\begin{subfigure}{0.2\linewidth}
		\includegraphics[width=\linewidth]{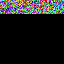}
		\caption{Encrypted Image}
        \label{fig:perfectdecrypt1a}
	\end{subfigure}
	\begin{subfigure}{0.2\linewidth}
		\includegraphics[width=\linewidth]{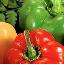}
		\caption{Decrypted Image}
        \label{fig:perfectdecrypt1b}
	\end{subfigure}
 \caption{Encrypted and Decrypted Image using K1 key for 64x64 pixel image}
\end{figure}

\begin{figure}
	\centering
	\begin{subfigure}{0.2\linewidth}
		\includegraphics[width=\linewidth]{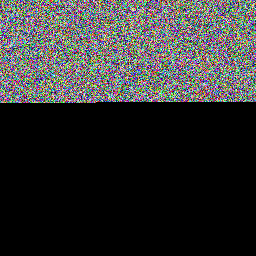}
		\caption{Encrypted Image}
        \label{fig:perfectdecrypt2a}
	\end{subfigure}
	\begin{subfigure}{0.2\linewidth}
		\includegraphics[width=\linewidth]{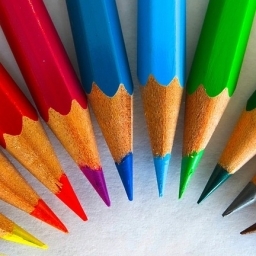}
		\caption{Decrypted Image}
        \label{fig:perfectdecrypt2b}
	\end{subfigure}
 \caption{Encrypted and Decrypted Image using K1 key for 256x256 pixel image}
\end{figure}

\begin{figure}
	\centering
	\begin{subfigure}{0.2\linewidth}
		\includegraphics[width=\linewidth]{visualized_encrypted_image_64.png}
		\caption{Encrypted Image}
        \label{fig:faileddecrypt1a}
	\end{subfigure}
	\begin{subfigure}{0.2\linewidth}
		\includegraphics[width=\linewidth]{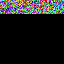}
		\caption{Decrypted Image}
        \label{fig:faileddecrypt1b}
	\end{subfigure}
 \caption{Encrypted using K1 and Decrypted using K2 key for 64x64 pixel image}
\end{figure}

\begin{figure}
	\centering
	\begin{subfigure}{0.2\linewidth}
		\includegraphics[width=\linewidth]{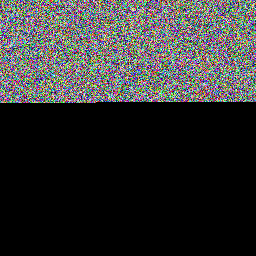}
		\caption{Encrypted Image}
        \label{fig:faileddecrypt2a}
	\end{subfigure}
	\begin{subfigure}{0.2\linewidth}
		\includegraphics[width=\linewidth]{visualized_encrypted_image_256_failed.png}
		\caption{Decrypted Image}
        \label{fig:faileddecrypt2b}
	\end{subfigure}
 \caption{Encrypted using K1 and Decrypted using K2 key for 256x256 pixel image}
\end{figure}

Key sensitivity is an essential property of cryptographic systems, ensuring that even a minor change in the key produces a vastly different ciphertext. This property guarantees that an attacker cannot derive useful information from closely related keys. \\
To demonstrate key sensitivity in our system, we conducted an experiment where we encrypted some color images with K1 and tried to decrypt the images with K2(where we altered one bit). The resulting ciphertexts were then compared to evaluate the difference. \\
For instance, using the keys, K1 = 000010010110011110101011111010111011101011111010111 and \\K2 = 100010010110011110101011111010111011101011111010111, the corresponding Hashed value of K1 = 8a49d097d696624218e1872935d9e3d2767bd9953d2e4a6b6946210966e5732c and the corresponding Hashed value of K2 = 200fa0c121fc9d2e2b7640445f05308ce671c68b29a1d2aae6311a392d468f5b\\
In figure \ref{fig:perfectdecrypt1a}, \ref{fig:perfectdecrypt1b}, \ref{fig:perfectdecrypt2a}, \ref{fig:perfectdecrypt2b}, \ref{fig:faileddecrypt1a}, \ref{fig:faileddecrypt1b}, \ref{fig:faileddecrypt2a}, \ref{fig:faileddecrypt2b} shows how high sensitivity our encryption system is. This ensures that even if an attacker gets close to the correct key, the decryption will still fail completely, thereby maintaining the security of the system.

\section{Discussion and Findings}
In our research, we have employed the Quantum Key Distribution Protocol E91 alongside some classical cryptographic methods. We encrypted and decrypted color images of various pixel densities, specifically 64x64, 128x128, 256x256, and 512x512. Our encryption process was independent of the resolution of the images.

We calculated and analyzed various results. We observed that the higher the pixel density of the images we encrypted, the more secure the encryption became. We calculated the entropy of the images and obtained remarkable results, with values $\approx 8$ on an 8-point scale. We also analyzed the differential attacks, such as NPCR and UACI, and found near-perfect results, as detailed in Section \ref{result}.

Furthermore, we determined the key generation rate and key length for different numbers of singlet states. Our findings indicate that around 25\% of the singlet states are used to generate the key, while the remaining singlet states are utilized to calculate the CHSH inequality, ensuring the detection of any eavesdroppers. We also measured the key generation time and rate, as well as the encryption and decryption times of our proposed system. As expected, larger pixel sizes required more time for encryption.

Lastly, we created histogram images to visually compare the differences between the encrypted and original images.

\section{Conclusions And Future Scope}
In our reseach, we proposed a robust and secure image data encryption system that integrates quantum key distribution (QKD) using the E91 protocol with classical cryptographic techniques, specifically the Advanced Encryption Standard (AES) and Secure Hash Algorithm. The system capitalizes on the inherent security advantages of quantum mechanics to ensure the confidentiality and integrity of transmitted data. Key contributions of this work include the utilization of the E91 protocol for generating cryptographic keys based on entangled photon pairs, the integration of QKD with AES to provide a strong encryption mechanism for securing image data, and the employment of deep learning-based steganography to embed encrypted images within cover images, further enhancing the security and privacy of data transmission. This method allows for the hidden transmission of sensitive information, making it less susceptible to detection. Our system demonstrates a practical application of combining quantum and classical cryptographic techniques with steganography, resulting in a highly secure image data encryption system. The proposed architecture is not only robust against potential eavesdropping but also ensures that the transmitted data remains confidential and intact.
While the proposed system showcases significant advancements in secure image data encryption, there are several avenues for future research and improvements. One area of future work involves investigating the incorporation of quantum error correction\cite{devitt2013quantum} techniques to further enhance the reliability of QKD in noisy environments\cite{nagata2017quantum}, thereby improving the robustness of key generation and distribution over longer distances. Additionally, exploring hybrid quantum-classical systems that combine the strengths of quantum and classical cryptography in more complex and diverse applications could lead to the development of protocols that seamlessly integrate with existing classical infrastructure. Enhancing the deep steganography component by incorporating more sophisticated neural network architectures and training techniques could improve the imperceptibility and robustness of the hidden data against various types of image processing attacks.

\section{Preliminary concepts}\label{precon}

\subsection{Quantum States}\label{qStates}
In quantum mechanics, a quantum state\cite{nielsen2010quantum} describes the state of a quantum system. For a qubit, has a two-dimensional state space, which can be expressed as a linear combination (or superposition) of the basis states $|0\rangle$ and $|1\rangle$.  Any arbitrary state of a qubit can be represented as: 
\begin{align}
    |\psi\rangle=\alpha|0\rangle + \beta|1\rangle
\end{align}
Here, $\alpha$ and $\beta$ are complex numbers. For the state $|\psi\rangle$ to be normalized (i.e., to be a unit vector), it must satisfy the condition:
\begin{align}
    \langle\psi|\psi\rangle\ = 1
\end{align}
This normalization condition implies:
\begin{align}
    |\alpha|^2 + |\beta|^2 = 1
\end{align}
This ensures that the total probability of finding the qubit in either state $|0\rangle$ or $|1\rangle$ is 1

\subsection{Entanglement}\label{entanglement}
When two qubits are entangled\cite{nielsen2010quantum}, their joint state cannot be expressed as a simple product of individual states. Instead, the state of the system is described by a single vector in the combined Hilbert space of the two qubits. So, a pure bipartite state $|\Psi\rangle_{AB}$ is entangled if it cannot be written as a product state $|\psi\rangle_A \otimes |\phi\rangle_B$ for any choices of states $|\psi\rangle_A$ and $|\phi\rangle_B$.\\
If we consider a density matrix of a pure state is $\rho$, where $\rho = |\psi\rangle\langle\psi|$. \\
A system is entangled if its density matrix cannot be written as a product of the density matrices of its subsystems.\\ 
Mathematically, this is expressed as:
\begin{align}
    \rho \neq \rho_A \otimes \rho_B
\end{align}
Where $\rho_A$ and $\rho_B$ are the density matrices of the subsystems A and B, respectively. \\
So, Bell states are defined as: 
\begin{align}
    \psi_{\pm}=\frac{1}{\sqrt{2}}|01\rangle \pm |10\rangle \\
    \phi_{\pm}=\frac{1}{\sqrt{2}}|00\rangle \pm |11\rangle
\end{align}

\subsection{E91 Protocol}\label{e91 protocol}
\begin{enumerate}
\item \textbf{Introduction:}\cite{e91main} In 1991, Artur Ekert developed a new approach to Quantum Key Distribution (QKD) by introducing the E91 protocol. This protocol ensures security through a Bell-like test to detect eavesdroppers (Eve).
\item \textbf{Entangled Particle Pairs:} A single source emits pairs of entangled particles, typically polarized photons. Each pair is described by a Bell state, specifically the singlet state:
\begin{align}
    \psi_{\pm}=\frac{1}{\sqrt{2}}|01\rangle \pm |10\rangle
\end{align}
This state implies that if one particle is measured to be in state $|0\rangle$, the other will be in state $|1\rangle$, and vice versa. The entangled particles are separated, with one particle sent to Alice and the other to Bob.\\
\item \textbf{Measurement Process:}\cite{pirandola2020advances} Alice and Bob each measure the polarization of their received photons. They randomly choose from three possible measurement bases, aligned according to the Clauser, Horne, Shimony, and Holt (CHSH) test.\\
\item \textbf{Measurement bases:}\label{measurement}\\ 
Alice's bases are:\\
\begin{align}
    a_1 = 0, a_2 = \frac{\pi}{4}, a_3 = \frac{\pi}{2}   
\end{align}
corresponding to the bases $Z$, $\frac{(X+Z)}{\sqrt{2}}$, and $X$, respectively.\\
Bob's bases are:\\
\begin{align}
    b_1 = \frac{\pi}{4}, b_2 = \frac{\pi}{2}, b_3 = \frac{3\pi}{2}
\end{align}
corresponding to the bases $\frac{(X+Z)}{\sqrt{2}}$, $X$, and $\frac{(X-Z)}{\sqrt{2}}$, respectively.
\item \textbf{Basis Discussion and Eavesdropper Detection:} Alice and Bob publicly discuss which measurement bases they used, without revealing their measurement outcomes. They use instances where they choose different bases to detect the presence of an eavesdropper.
\item \textbf{CHSH Test:}\label{CHSH} The CHSH inequality helps detect eavesdropping by checking the correlation between Alice’s and Bob’s measurements:
\begin{align}
    E = \langle a_1b_1 \rangle - \langle a_1b_3 \rangle + \langle a_3b_1 \rangle + \langle a_3b_3 \rangle
\end{align}
$\langle a_ib_j \rangle$ represents the expectation value when Alice measures using basis $a_i$ and Bob using basis $b_j$. If $-2 \le E \le 2$, it indicates potential eavesdropping or issues with the measurement devices. In an ideal, eavesdropper-free system, the expected value is $-2\sqrt{2}$, indicating maximal violation of the CHSH inequality.
\item \textbf{State of Entangled Photons:} Entangled photons passing through a depolarizing channel can be described by the isotropic mixed state\cite{nielsen2010quantum}: 
\begin{align}
    \rho_{\psi} = p|\psi\rangle\langle\psi| + \frac{(1-p)}{4}I
\end{align}
Here, $p$ is the probability of maintaining the pure entangled state, and $I$ is the identity matrix. The CHSH test shows maximal violation $-2\sqrt{2}$ when $p = 1$, meaning the channel is free from eavesdropping. 
\item \textbf{Key Generation:} When Alice and Bob detect maximal violation of the CHSH test, they are confident their data is secure and not influenced by an eavesdropper. They use the instances where they chose the same measurement bases to generate a shared private key, utilizing their perfectly anti-correlated measurement results.
\item \textbf{Essential Features of E91:} The E91 protocol capitalizes on the non-local nature of entangled states, which ensures security against eavesdropping. Eve’s intervention would introduce detectable changes in the entanglement correlations, affecting the non-locality and alerting Alice and Bob to her presence. This protocol, therefore, provides a robust framework for secure quantum communication.
\end{enumerate}

\subsection{Quantum Logic Gates}\label{quantum logic gate}
The logic flow in a quantum circuit\cite{sagawa2011fundamentals} involves the arrival of an input qubit state (which is a quantum state) that is operated upon by a unitary operator and an output qubit state is obtained, following the laws or postulates of quantum mechanics.\\
If $U$ is a unitary operator representing a quantum gate, and $|\psi\rangle$ is the input qubit state, the output qubit state $|\psi'\rangle$ is given by:
\begin{align}
    |\psi'\rangle = U|\psi\rangle
\end{align}

Unitary operators have the property that satisfies $U^{\dag} U = UU^{\dag} = I$, where $U^{\dag}$ is the Hermitian Conjugate of $U$, and $I$ is the identity matrix. This property ensures that quantum gates are reversible and preserve the total probability. \\
Some important quantum logic gates are given below which we need to implement the QKD protocol.
\begin{enumerate}
    \item \textbf{Pauli-$X$ Gate:} The Pauli-$X$ gate flips the state of a qubit, analogous to the classical NOT gate.\\
    It is represented as:
    \begin{align}
        X = 
        \begin{pmatrix}
        0 & 1\\
        1 & 0
        \end{pmatrix}
    \end{align}

    \item \textbf{Hadamard Gate:} The Hadamard gate creates superposition states from basis states. It is essential in algorithms that rely on quantum parallelism.\\ 
    The Hadamard gate is represented as:
    \begin{align}
        H = \frac{1}{\sqrt{2}}
        \begin{pmatrix}
        1 & 1\\
        1 & -1
        \end{pmatrix}
    \end{align}
    \item \textbf{Controlled-NOT (C-NOT) Gate:} The C-NOT gate is a two-qubit gate that flips the state of the target qubit if the control qubit is in the state $|1\rangle$.\\ 
    It is represented as:
    \begin{align}
        CNOT = 
        \begin{pmatrix}
        1 & 0 & 0 & 0\\
        0 & 1 & 0 & 0\\
        0 & 0 & 0 & 1\\
        0 & 0 & 1 & 0\\
        \end{pmatrix}
    \end{align}
    \item \textbf{Phase Gate:} The Phase gate (S gate) introduces a phase shift of $\frac{\pi}{2}$.\\
    It is represented as:
    \begin{align}
        S = 
        \begin{pmatrix}
        1 & 0\\
        0 & i
        \end{pmatrix}
    \end{align}
    \item \textbf{$T$ Gate:} The $T$ gate, or $\frac{\pi}{8}$ gate, introduces a phase shift of $\frac{\pi}{4}$.\\
    It is represented as:
    \begin{align}
        T = 
        \begin{pmatrix}
        1 & 0\\
        0 & e^\frac{i\pi}{4}
        \end{pmatrix}
    \end{align}
\end{enumerate}

\subsection{Advanced Encryption Standard (AES)}\label{AES}
The Advanced Encryption Standard (AES) is a symmetric encryption algorithm widely used across various applications for securing sensitive data. Developed by Joan Daemen and Vincent Rijmen, AES was adopted as a standard by the National Institute of Standards and Technology (NIST) in 2001. It is known for its efficiency, security, and flexibility.\\
AES utilizes a design principle called a substitution-permutation network, making it efficient for both software and hardware implementations. It is a specific variant of the Rijndael algorithm, featuring a fixed block size of 128 bits and key sizes of 128, 192, or 256 bits. In comparison, the Rijndael algorithm itself allows for block and key sizes that can be any multiple of 32 bits, ranging from 128 to 256 bits.\\
\textbf{Description\cite{aesnist} of the AES algorithm:} 
\begin{enumerate}
    \item KeyExpansion: Round keys are derived from the cipher key using the AES key schedule. AES requires a separate 128-bit round key block for each round plus one more.
    \item Initial round key addition:
    \begin{enumerate}
        \item AddRoundKey - each byte of the state is combined with a byte of the round key using bitwise xor.
    \end{enumerate}
    \item 9, 11 or 13 rounds:
    \begin{enumerate}
        \item SubBytes – A non-linear substitution step where each byte is replaced with another according to Rijndael S-box.
        \item ShiftRows – A transposition step where the last three rows of the state are shifted cyclically over a certain number of steps.
        \item MixColumns – A linear mixing operation that operates on the columns of the state, combining the four bytes in each column.
        \item AddRoundKey
    \end{enumerate}
    \item Final round (making 10, 12 or 14 rounds in total):
    \begin{enumerate}
        \item SubBytes
        \item ShiftRows
        \item AddRoundKey
    \end{enumerate}
\end{enumerate}
\textbf{SubBytes() Transformation:} In the SubBytes step, each byte $a_{i,j}$ in the state array is replaced by a SubByte $S(a_{i,j})$ using an 8-bit substitution box. This S-box which is invertible, is constructed by composing two transformations:
\begin{enumerate}
    \item Take the multiplicative inverse in the finite field $GF(2^8)$; the element $\{00\}$ is mapped to itself.
    \item Apply the following affine transformation (over $GF(2)$):
    \begin{align}
        b_i = b_i \oplus b_{(i+4)mod8} \oplus b_{(i+5)mod8} \oplus b_{(i+6)mod8} \oplus b_{(i+7)mod8} \oplus c_i
    \end{align}
    for $0\le i < 8$, where $b_i$ is the $i^{th}$ bit of the byte, and $c_i$ is the $i^{th}$ bit of a byte $c$ with the value $\{63\}$ or $\{01100011\}$. Here and elsewhere, a prime on a variable (e.g., $b'$) indicates that the variable is to be updated with the value on the right.
\end{enumerate}


\textbf{ShiftRows() Transformation:} In the ShiftRows() transformation, the bytes in the last three rows of the State are cyclically shifted over different numbers of bytes (offsets). The first row, $r=0$, is not shifted.\\
Specifically, the ShiftRows() transformation proceeds as follows:
\begin{align}
    S_{r,c} = S_{r, (c + shift(r, Nb)) mod Nb} \ for \ 0 < r < 4 \ and \ 0 \le c < Nb
\end{align}

where the shift value $shift(r,Nb)$ depends on the row number, $r$, as follows (Number of Columns, for this standand $Nb=4$):
\begin{align}
    shift(1,4)=1; \ shift(2, 4) = 2; \ shift(3, 4) = 3
\end{align}


\textbf{MixColumns() Transformation:} The MixColumns() transformation operates on the State column-by-column, treating each column as a four-term polynomial. The columns are considered as polynomials over $GF(2^8)$ and multiplied modulo $x^4 + 1$ with a fixed polynomial $a(x)$, given by
\begin{align}
    a(x) = \{03\}x^3 + \{01\}x^2 + \{01\}x + \{02\}
\end{align}

This can be written as a matrix multiplication.
\begin{align}
    s'(x)=a(x) \oplus s(x):
    \begin{bmatrix}
        s'_{0,c} \\  
        s'_{1,c} \\
        s'_{2,c} \\
        s'_{3,c} 
    \end{bmatrix}
    =
    \begin{bmatrix}
        02 & 03 & 01 & 01 \\
        01 & 02 & 03 & 01 \\
        01 & 01 & 02 & 03 \\
        03 & 01 & 01 & 02 
    \end{bmatrix}
    \begin{bmatrix}
        s_{0,c} \\  
        s_{1,c} \\
        s_{2,c} \\
        s_{3,c}
    \end{bmatrix}
    \ for \ 0 \le c < Nb
\end{align}


\textbf{AddRoundKey() Transformation:} In the AddRoundKey() transformation, a Round Key is added to the State by a simple bitwise XOR operation.Each Round Key consists of Nb words from the key schedule. Those Nb words are each added into the columns of the State, such that
\begin{align}
    \begin{bmatrix}
        s'_{0,c}, s'_{1,c}, s'_{2,c}, s'_{3,c} 
    \end{bmatrix}
    =
    \begin{bmatrix}
        s'_{0,c}, s'_{1,c}, s'_{2,c}, s'_{3,c}   
    \end{bmatrix}
    \oplus
    \begin{bmatrix}
        W_{round*Nb+c}
    \end{bmatrix}
    \ for \ 0 \le c < Nb
\end{align}
where $[w_i]$ is the key schedule words and round is a value in the range $0 \le round \le Nr \ (Number \  of \ rounds = 10, 12, 14)$.

\subsection{Secure Hash Algorithm (SHA)}\label{SHA}
The Secure Hash Algorithm (SHA) family is a set of cryptographic hash functions designed to provide strong data integrity and security. SHA was developed by the NIST and has various versions, including SHA-1, SHA-2, and SHA-3. Each version of SHA processes\cite{penard2008secure} data in fixed-size blocks and produces a fixed-size hash value, making it ideal for various security applications such as digital signatures, message integrity verification, and password hashing.

\subsection{Steps of SHA-256:}\label{subsubsec1}

Here, we will focus on the SHA-2 family, specifically SHA-256, which is widely used due to its balance of security and performance. SHA-256 operates through several main steps: message padding, parsing, initialization, message scheduling, and the compression function.

\begin{enumerate}
    \item \textbf{Message Padding:} The message to be hashed is padded to ensure its length is a multiple of 512 bits (64 bytes). Padding is done as follows:
    \begin{enumerate}
        \item Append a '1' bit: Add a single '1' bit to the end of the message. 
        \item Append '0' bits: Add '0' bits until the length of the message is 64 bits less than a multiple of 512.
        \item Append the length: Append the 64-bit representation of the original message length (in bits) to the end of the padded message.
    \end{enumerate}

    \item \textbf{Parsing the Message:} The padded message is divided into 512-bit (64-byte) blocks. Each block is further divided into sixteen 32-bit words:

    \begin{align}
        M = {M_1, M_2, M3,.....,M_N}
    \end{align}
    
    \item \textbf{Initialization of Hash Values:} SHA-256 uses eight initial hash values, each 32 bits in length. These values are derived from the fractional parts of the square roots of the first eight prime numbers:
    
    \begin{align}
        \begin{split}
            H_0 = 0x6a09e667, H_1 = 0xbb67ae85, H_2 = 0x3c6ef372, H_3 = 0xa54ff53a, \\
            H_4 = 0x510e527f, H_5 = 0x9b05688c, H_6 = 0x1f83d9ab, H_7 = 0x5be0cd19
        \end{split}
    \end{align}
    
    \item \textbf{Message Schedule:} For each 512-bit block, a message schedule consisting of 64 32-bit words is created. The first 16 words are the original block words, and the remaining 48 words are generated using the following formula:
    
    \begin{align}
        W_t=\begin{cases}
            M_t & \text{for $0 \le t \le 15$} \\
            \sigma_1(W_{t-2}) + W_{t-7} + \sigma_0(W_{t-15}) + W_{t-16} & \text{for $16 \le t \le 63$}
        \end{cases}
    \end{align}
    where,
    $\sigma_0(x) = (x \gg 7) \oplus (x \gg 18) \oplus (x \gg 3)$  and $\sigma_1(x) = (x \gg 17) \oplus (x \gg 19) \oplus (x \gg 10)$

    \item \textbf{Compression Function:} The core of SHA-256 is the compression function, which processes each 512-bit block with the initialized hash values. The compression function involves the following steps:
    \begin{enumerate}
        \item Initialize Working Variables: Initialize eight working variables $a, b, c, d, e, f, g, h$ with the current hash values $H_0, H_1, H_2, H_3, H_4, H_5, H_6, H_7$.
        \item Main Loop: Iterate through each of the 64 rounds using:
        \begin{align}
            T_1 = h + \sum_1(e) + Ch(e, f,g) + K_t + W_t \\
            T_2 = \sum_0(a) + Maj(a, b, c) \\
            h = g, g = f, f = e, e = d + T_i \\
            d = c, c = b, b = a, a = T_1 + T_2
        \end{align}
        where,
        \begin{align}
            \sum_0(x) = (x \gg 2) \oplus (x \gg 13) \oplus (x \gg 22) \\
            \sum_1(x) = (x \gg 6) \oplus (x \gg 11) \oplus (x \gg 25) \\
            Ch(x, y, z) = (x \wedge y) \oplus (\neg x \wedge z) \\
            Maj(x, y, z) = (x \wedge y) \oplus (x \wedge z) \oplus (y \wedge z)
        \end{align}
        
        \item Update Hash Values: After completing the rounds for each block, update the hash values:
        \begin{align}
            \begin{split}
                H_0 = H_0 + a, H_1 = H_1 + b, H_2 = H_2 + c, H_3 = H_3 + d \\
                H_4 = H_4 + e, H_5 = H_5 + f, H_6 = H_6 + g, H_7 = H_7 + h
            \end{split}
        \end{align}
    \end{enumerate}

    \item \textbf{Final Hash Value:} After processing all blocks, the final hash value is the concatenation of the updated hash values:
    \begin{align}\label{hasheqn}
        \text{Hash} = H_0 \parallel H_1 \parallel H_2 \parallel H_3 \parallel H_4 \parallel H_5 \parallel H_6 \parallel H_7
    \end{align}
    This 256-bit hash value serves as the output of the SHA-256 algorithm.
\end{enumerate}

\subsection{Steganography}\label{stega}
Steganography\cite{cheddad2010digital} is the practice of concealing information within another non-secret medium to prevent detection. The word itself derives from the Greek words "steganos," meaning covered, and "graphein," meaning writing.

\subsubsection{LSB-based steganography:}\label{LSB} Most commonly used technique to hide image data within image data. LSB-based steganography\cite{chandramouli2001analysis} modifies the last bit of a pixel's RGB value to embed a single bit of the secret data. 
\begin{align}
    Pixel(i, j) = (R_{i, j}, G_{i, j}, B_{i, j}) \  & \ \text{Original Pixel Value.}\\
    Pixel'(i, j) = (R'_{i, j}, G'_{i, j}, B'_{i, j}) \ & \ \text{Modified Pixel Value after altering last bit.}
\end{align}

\subsubsection{Deep Steganography:}\label{deepstega} Deep Learning based Steganography\cite{baluja2017hiding} has the potential to create lossless images. LSB based steganography will always have some pixel altered but deep steganography uses Convolution Neural Network to hide images within images. This method distributes the secret image's representation across all available bits of the cover image. This disperses the hidden information, making it less detectable.
\begin{align}
    L(c, c', s, s') = \parallel c - c' \parallel + \beta \parallel s - s' \parallel
\end{align}
Here, $c$ and $s$ are the cover and secret images, while $c'$ and $s'$ are the reconstructed versions. $L$ is the loss function, and $\beta$ controls the trade-off.

\section{Acknowledgement}
M.R.C. Mahdy acknowledges the support of NSU internal grant and CTRGC grant 2023-24.

\section{Competing Interests}
The author declares that there is no competing financial interests.



\printcredits

\bibliographystyle{model1-num-names}

\bibliography{cas-refs}

\end{document}